\documentclass{article}

\usepackage{subcaption,graphicx,amsmath,amsfonts,fullpage}

\begin{document}

\title{A neural network system for transformation of regional cuisine style} 
\author{Masahiro Kazama,$^{1}$ Minami Sugimoto,$^{2}$ Chizuru Hosokawa,$^{2}$ Keisuke Matsushima,$^{3}$\\  Lav R. Varshney,$^{4}$ and Yoshiki Ishikawa$^{1}$\thanks{ishikun@gmail.com} \vspace{3mm}\\
$^{1}$Habitech Inc., Tokyo, Japan\\
$^{2}$Department of Public Health, Cancer Scan Inc., Tokyo, Japan\\
$^{3}$Keisuke Matsushima, Nice, France\\
$^{4}$University of Illinois at Urbana-Champaign, Illinois, USA }
\date{}

\maketitle

\begin{abstract}
We propose a novel system which can transform a recipe into any selected regional style (e.g., Japanese, Mediterranean, or Italian). This system has two characteristics. First the system can identify the degree of regional cuisine style mixture of any selected recipe and visualize such regional cuisine style mixtures using barycentric Newton diagrams. Second, the system can suggest ingredient substitutions through an extended word2vec model, such that a recipe becomes more authentic for any selected regional cuisine style. Drawing on a large number of recipes from Yummly, an example shows how the proposed system can transform a traditional Japanese recipe, Sukiyaki, into French style.

\vspace{3mm}
\end{abstract}

\section{Introduction}

With growing diversity in personal food preference and regional cuisine style, personalized information systems that can transform a recipe into any selected regional cuisine style that a user might prefer would help food companies and professional chefs create new recipes.

To achieve this goal, there are two significant challenges: 1) identifying the degree of regional cuisine style mixture of any selected recipe; and 2) developing an algorithm that shifts a recipe into any selected regional cuisine style. 

As to the former challenge, with growing globalization and economic development, it is becoming difficult to identify a recipe’s regional cuisine style with specific traditional styles since regional cuisine patterns have been changing and converging in many countries throughout Asia, Europe, and elsewhere \cite{khoury2014increasing}. Regarding the latter challenge, to the best of our knowledge, little attention has been paid to developing algorithms which transform a recipe’s regional cuisine style into any selected regional cuisine pattern, cf. \cite{pinel2014using,pinel2014substitution}. Previous studies have focused on developing an algorithm which suggests replaceable ingredients based on cooking action \cite{shidochi2009finding}, degree of similarity among ingredient \cite{nozawa}, ingredient network \cite{teng2012recipe}, degree of typicality of ingredient \cite{yokoi2015typicality}, and flavor (foodpairing.com).

The aim of this study is to propose a novel data-driven system for transformation of regional cuisine style. This system has two characteristics. First, we propose a new method for identifying a recipe’s regional cuisine style mixture by calculating the contribution of each ingredient to certain regional cuisine patterns, such as Mediterranean, French, or Japanese, by drawing on ingredient prevalence data from large recipe repositories. Also the system visualizes a recipe’s regional cuisine style mixture in two-dimensional space under barycentric coordinates using what we call a {\it Newton diagram}. Second, the system transforms a recipe’s regional cuisine pattern into any selected regional style by recommending replaceable ingredients in existing recipes.

As an example of this proposed system, we transform a traditional Japanese recipe, Sukiyaki, into French style.

\section{Architecture of transformation system}
Figure \ref{fig1} shows the overall architecture of the transformation system, which consists of two steps: 1) identification and visualization of a recipe’s regional cuisine style mixture; and 2) algorithm which transforms a given recipe into any selected regional/country style. Details of the steps are described as follows.

\begin{figure}
\begin{center}
\includegraphics[width=15cm]{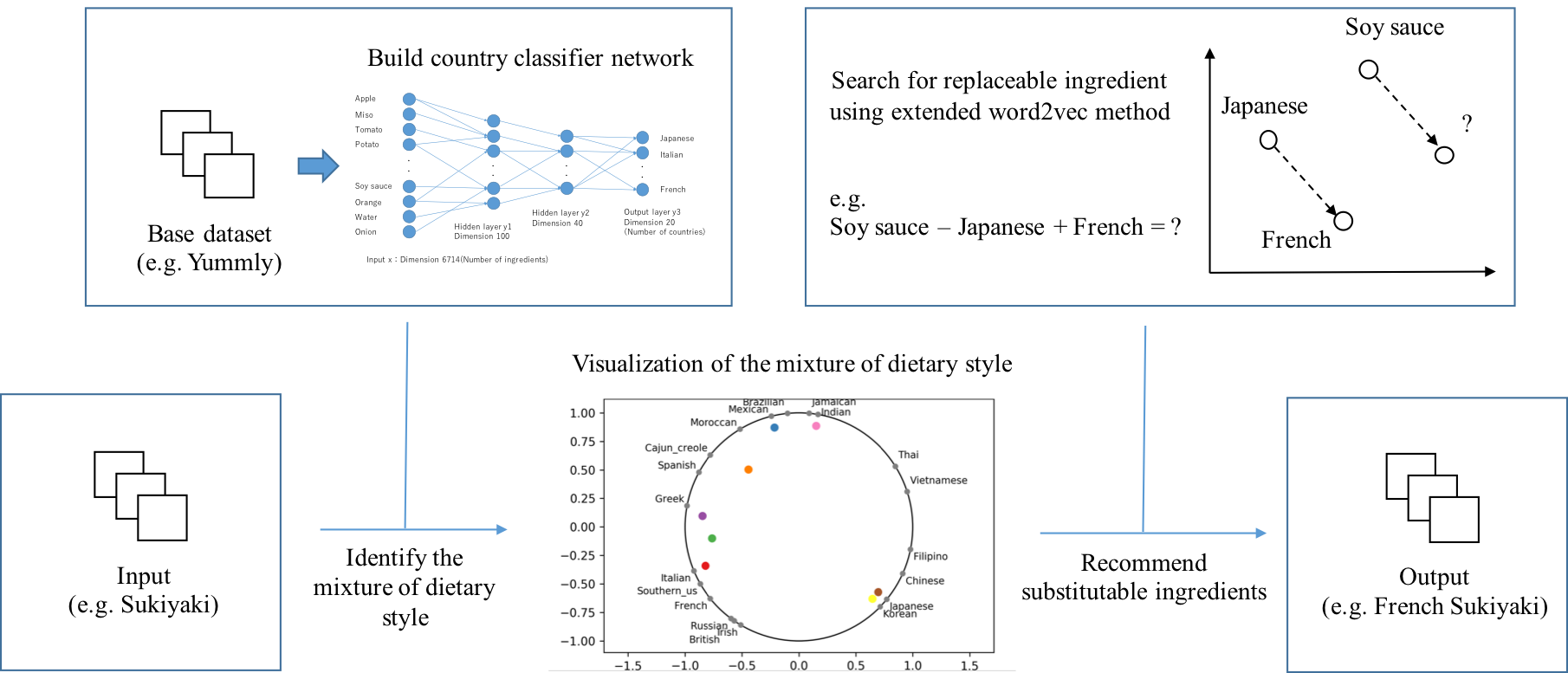}
\end{center}
\caption{ Architecture of transformation system which transform a given recipe into any selected country/region style}\label{fig1}
\label{fig1}
\end{figure}

\begin{figure}
\begin{center}
\includegraphics[width=15cm]{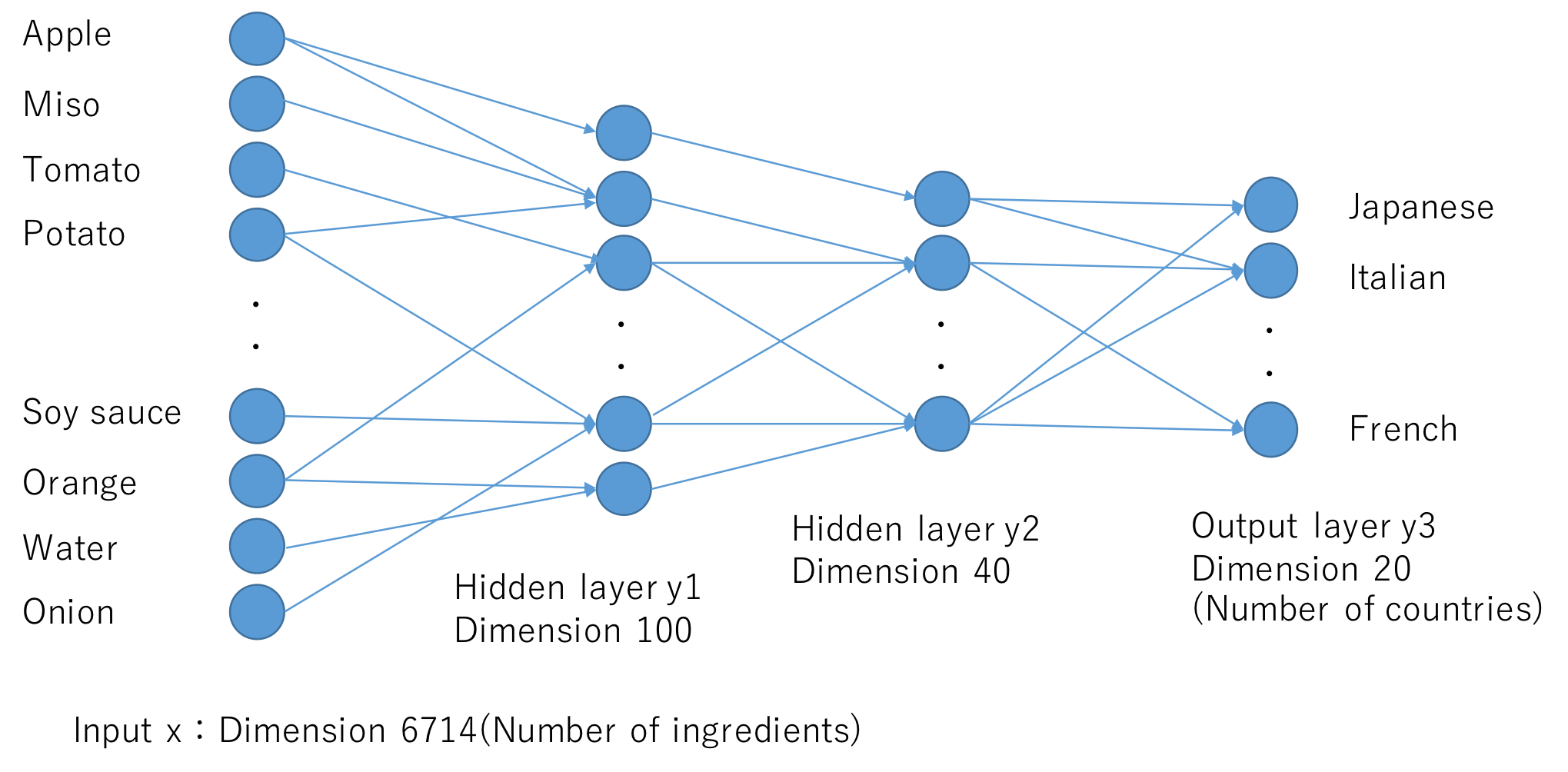}
\end{center}
\caption{ Neural network model for predicting regional cuisine from list of ingredients.}
\label{fig2}
\end{figure}

\subsection{Step 1: Identification and visualization of a recipe's regional cuisine style mixture}
Using a neural network method as detailed below, we identify a recipe's regional cuisine style. The neural network model was constructed as shown in Figure \ref{fig2}. The number of layers and dimension of each layer are also shown in Figure \ref{fig2}. 

When we enter a recipe, this model classifies which country or regional cuisine the recipe belongs to. The input is a vector with the dimension of the total number of ingredients included in the dataset, and only the indices of ingredients contained in the input recipe are 1, otherwise they are 0. 

There are two hidden layers. Therefore, this model can consider a combination of ingredients to predict the country probability. Dropout is also used for the hidden layer, randomly (20\%) setting the value of the node to 0. So a robust network is constructed. The final layer’s dimension is the number of countries, here 20 countries. In the final layer, we convert it to a probability value using the softmax function, which represents the probability that the recipe belongs to that country. 
ADAM \cite{kingma2014adam} was used as an optimization technique. The number of epochs in training was 200. These network structure and parameters were chosen after preliminary experiments so that the neural network could perform the country classification task as efficiently as possible.

\begin{table}[htbp]
\begin{center}
\caption{Statistics of Yummly dataset and some recipe examples.}
\begin{tabular}{cc}
\begin{minipage}{0.5\hsize}
\begin{center}
\begin{tabular}{|l|c|r|}
\hline
Country &Recipes & Ingredients  \\ \hline \hline
Italian	&	7838	&	2929	\\ \hline
Mexican	&	6438	&	2684	\\ \hline
Southern US	&	4320	&	2462	\\ \hline
Indian	&	3003	&	1664	\\ \hline
Chinese	&	2673	&	1792	\\ \hline
French	&	2646	&	2102	\\ \hline
Cajun Creole	&	1546	&	1576	\\ \hline
Thai	&	1539	&	1376	\\ \hline
Japanese	&	1423	&	1439	\\ \hline
Greek	&	1175	&	1198	\\ 
\hline
\end{tabular}
\end{center}
\end{minipage}

\begin{minipage}{0.5\hsize}
\begin{tabular}{|l|c|r|}
\hline
Country &Recipes & Ingredients  \\ \hline \hline
Spanish	&	989	&	1263	\\ \hline
Korean	&	830	&	898	\\ \hline
Vietnamese	&	825	&	1108	\\ \hline
Moroccan	&	821	&	974	\\ \hline
British	&	804	&	1166	\\ \hline
Filipino	&	755	&	947	\\ \hline
Irish	&	667	&	999	\\ \hline
Jamaican	&	526	&	877	\\ \hline
Russian	&	489	&	872	\\ \hline
Brazilian	&	467	&	853	\\ \hline \hline
ALL	&	39774	&	6714	\\ 
\hline
\end{tabular}
\end{minipage}

\end{tabular}

\begin{minipage}{0.9\hsize}
\begin{tabular}{|l|c|l|}
\hline
RecipeID &Country& Ingredients  \\ \hline \hline
34466 & British & greek yogurt, lemon curd, confectioners sugar, raspberries\\ \hline
44500  & Indian & chili, mayonaise, chopped onion, cider vinegar, fresh mint, cilantro leaves\\ \hline
38233  & Thai & sugar, chicken thighs, cooking oil, fish sauce, garlic, black pepper\\ \hline
\end{tabular}
\end{minipage}

\label{table1}
\end{center}
\end{table}

\begin{table}[htbp]
\begin{center}
\caption{Example of ingredient classification by the neural network. Three top countries are listed with the probability that the ingredient are classified into.}
\begin{tabular}{|c|c|r|r|r|}
\hline
 Ingredient & Top1 & Top2 & Top3\\ \hline \hline
Onions  & French & Italian & Mexican \\ 
  & 0.130 & 0.126 & 0.126 \\ \hline
Soy sauce  & Japanese & Chinese & Filipino \\ 
  & 0.246 & 0.233 & 0.122 \\ \hline
 Mirin & Japanese & French & Korean \\ 
 & 0.890 & 0.040 &0.020 \\ 
\hline
\end{tabular}
\label{table1a}
\end{center}
\end{table}

\begin{figure}[h!]
\begin{center}
\includegraphics[width=15cm]{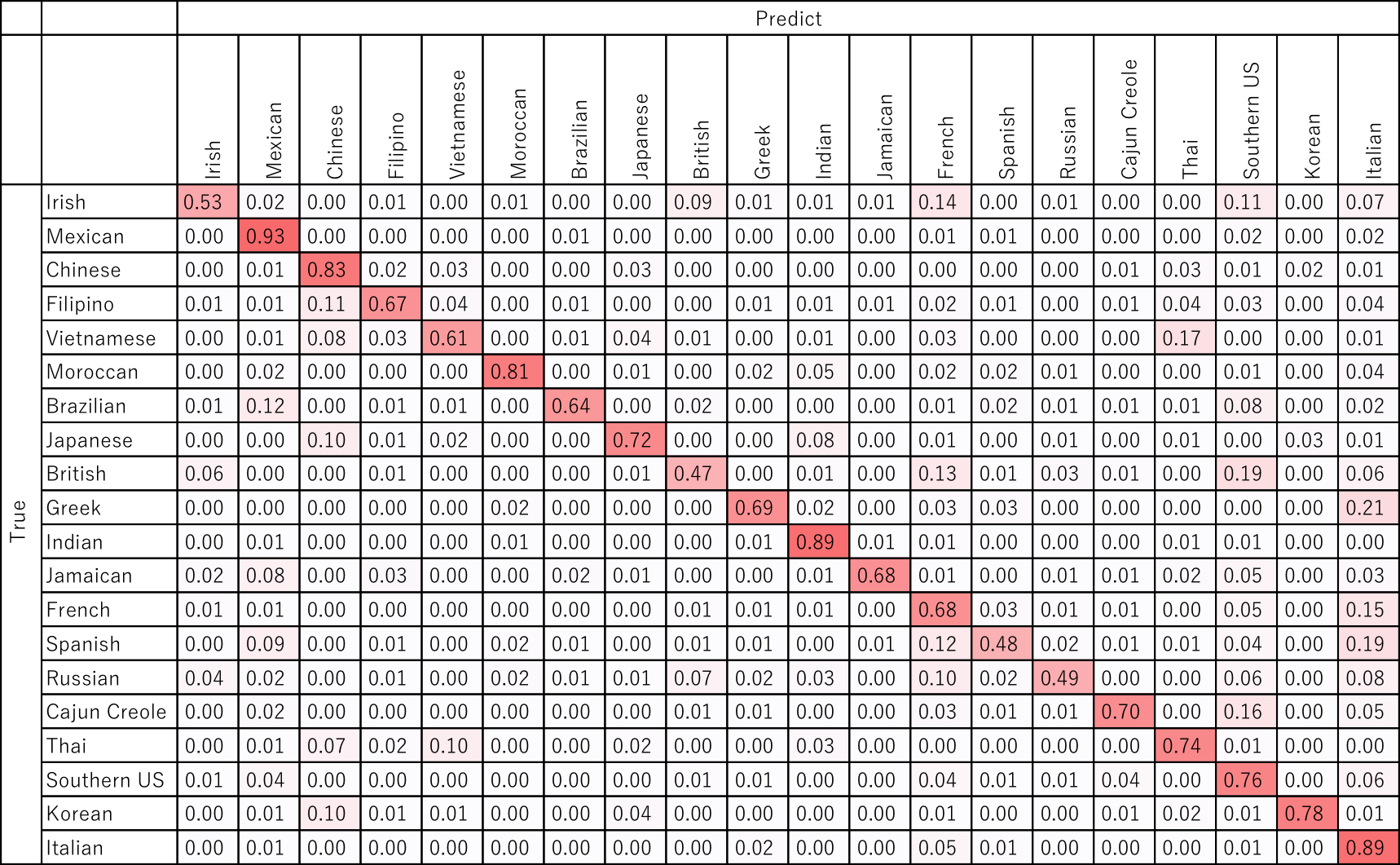}% This is a *.jpg file
\end{center}
\caption{Confusion matrix of neural network classiffication. }\label{fig.conf}
\end{figure}

In this study, we used a labeled corpus of Yummly recipes to train this neural network. Yummly dataset has 39774 recipes from the 20 countries as shown in Table \ref{table1}. Each recipe has the ingredients and country information. Firstly, we randomly divided the data set into 80\% for training the neural network and 20\% for testing how precisely it can classify. The final neural network achieved a classification accuracy of 79\% on the test set. Figure \ref{fig.conf} shows the confusion matrix of the neural network classifficaiton. Table \ref{table1a} shows the examples of ingredient classification results. Common ingredients, onions for example, that appear in many regional recipes are assigned to all countries with low probability. On the other hands some ingredients that appear only in specific country are assigned to the country with high probability. For example mirin that is a seasoning commonly used in Japan is classified into Japan with high probability.

\begin{figure}[h!]
\begin{center}
\includegraphics[width=15cm]{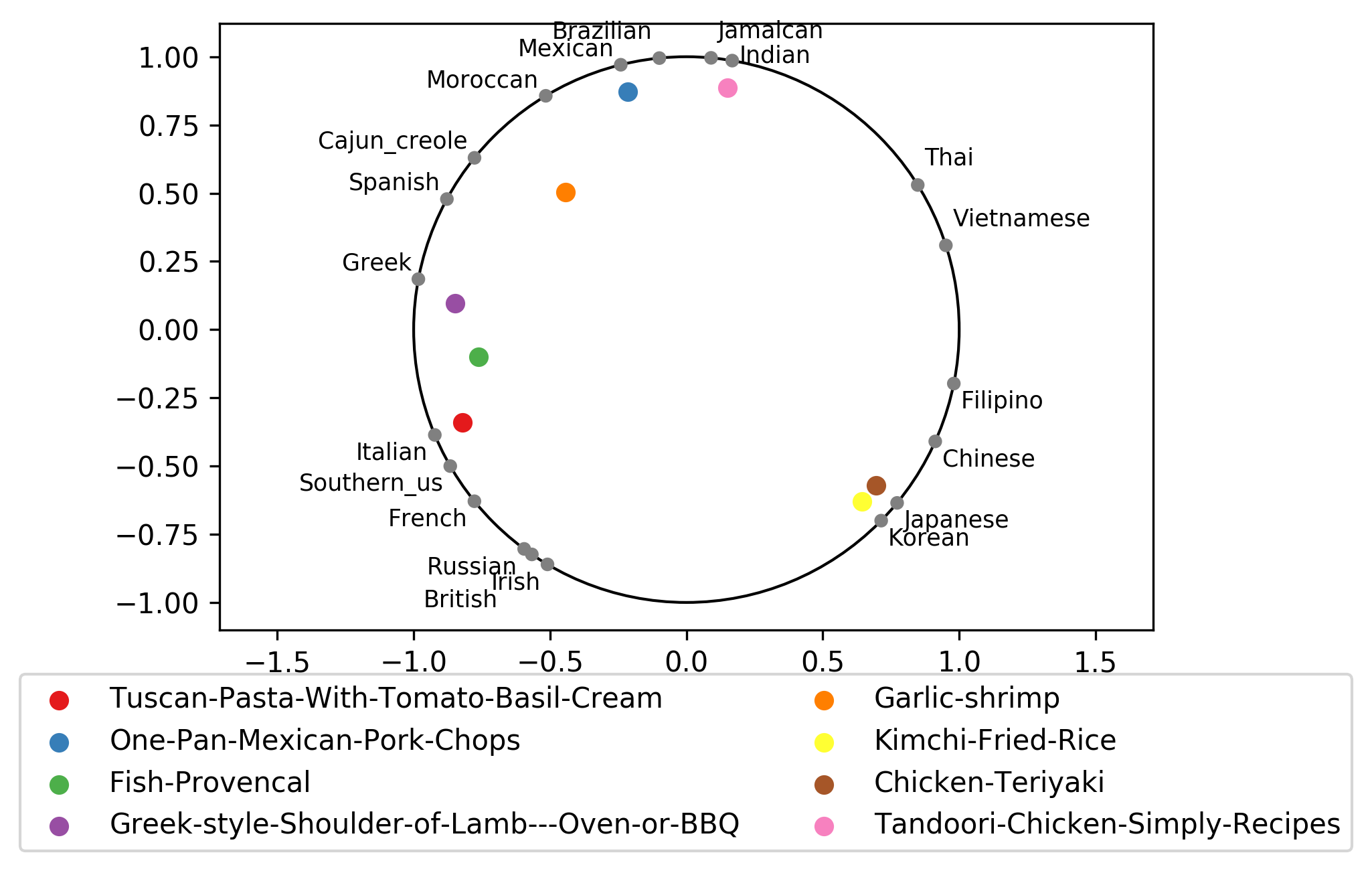}% This is a *.jpg file
\end{center}
\caption{ Newton diagram: visualization of probability that the recipe belongs to the several regional cuisine style. Countries are placed by spectral drawing.
%We pick 8 recipes and 8 countries as an example.
}\label{fig3}
\end{figure}

By using the probability values that emerge from the activation function in the neural network, rather than just the final classification, we can draw a barycentric Newton diagram, as shown in Figure \ref{fig3}.  The basic idea of the visualization, drawing on Isaac Newton’s visualization of the color spectrum \cite{newton1704}, is to express a mixture in terms of its constituents as represented in barycentric coordinates.  This visualization allows an intuitive interpretation of which country a recipe belongs to. If the probability of Japanese is high, the recipe is mapped near the Japanese. The countries on the Newton diagram are placed by spectral graph drawing \cite{koren2003spectral}, so that similar countries are placed nearby on the circle. The calculation is as follows.
First we define the adjacency matrix $W$ as the similarity between two countries. 
The similarity between country $i$ and $j$ is calculated by cosine similarity of county $i$ vector and $j$ vector. These vector are defined in next section. $W_{ij} = sim(vec_i, vec_j)$.
The degree matrix $D$ is a diagonal matrix where $D_{ii} = \sum_{j} W_{ij}$. 
Next we calculate the eigendecomposition of $D^{-1}W$. The second and third smallest eingenvalues and corresponded eingevectors are used for placing the countries. Eigenvectors are normalized so as to place the countries on the circle. 

\subsection{Step 2: Transformation algorithm for transforming regional cuisine style}
If you want to change a given recipe into a recipe having high probability of a specific country by just changing one ingredient, which ingredient should be alternatively used?

When we change the one ingredient $x_i$ in the recipe to ingredient $x_j$, the probability value of country likelihood can be calculated by using the above neural network model. If we want to change the recipe to have high probability of a specific country $c$, we can find ingredient $x_j$ that maximizes the following probability. $P(C=c|r - x_i + x_j)$
%\begin{equation}
%P(C=c|r - x_i + x_j)\label{eq:01}
%\end{equation}
where $r$ is the recipe.
However, with this method, regardless of the ingredient $x_i$, only specific ingredients having a high probability of country $c$ are always selected. In this system, we want to select ingredients that are similar to ingredient $x_i$ and have a high probability of country $c$. Therefore, we propose a method of extending word2vec as a method of finding ingredients resembling ingredient $x_i$.

\begin{figure}[h!]
\begin{center}
\includegraphics[width=15cm]{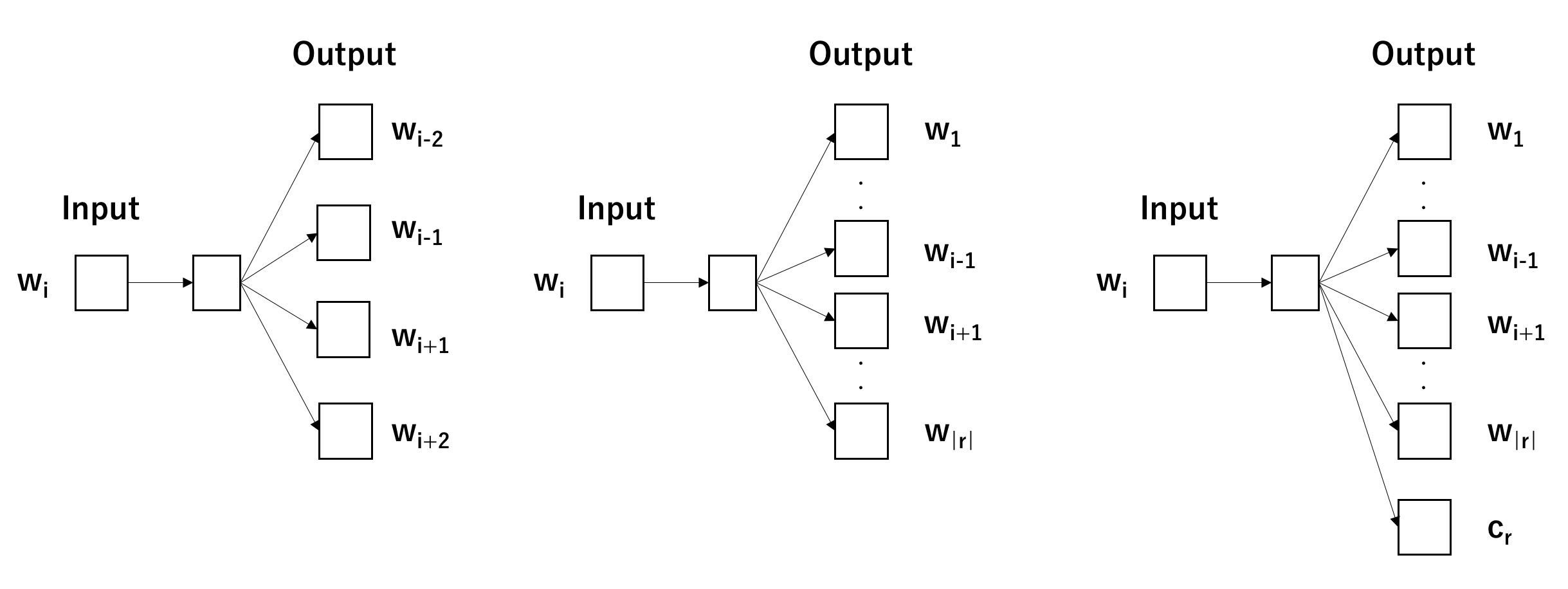}% This is a *.jpg file
\end{center}
\caption{The word2vec (skip-gram) architecture. The left panel is the traditional word2vec with window size $n=2$. The middle panel is the word2vec for recipe data. The right panel is the word2vec for recipe data with country information. 
%We pick 8 recipes and 8 countries as an example.
}\label{word2vec_arch}
\end{figure}

Word2vec is a technique proposed in the field of natural language processing \cite{mikolov2013distributed}. As the name implies, it is a method to vectorize words, and similar words are represented by similar vectors. To train word2vec, skip-gram model is used. In the skip-gram model, the objective is to learn word vector representations that can predict the nearby words. The objective function is 

\begin{equation}
\sum_{d \in D} \sum_{w_i \in d} \sum_{-n \leq j \leq n, j \neq 0} \log P(w_{i + j}|w_i) \label{eq:02}
\end{equation}
where $D$ is the set of documents, $d$ is a document, $w_i$ is a word, and $n$ is the window size. This model predicts the $n$ words before and after the input word, as described in left side of Figure \ref{word2vec_arch}. The objective function is to maximize the likelihood of the prediction of the surrounding word $w_{i+j}$ given the center word $w_i$.
The probability is 
\begin{equation}
P(w_j|w_i) = \frac{\exp(v_{w_i}^Tv_{w_j}^{'})}{\sum_{w \in W} \exp(v_{w_i}^Tv_w^{'})}
\end{equation}
where $v_w \in \mathbb{R}^K$ is an input vector of word $w$, $v^{'}_w \in \mathbb{R}^K$ is an output vector of word $w$, $K$ is the dimension of the vector, and $W$ is the set of all words. To optimize this objective function, hierarchical softmax or negative sampling method \cite{mikolov2013distributed} are used. After that we get the vectors of words and we can calculate analogies by using the vectors. For example, the analogy of  ``King - Man + Women = ?" yields ``Queen" by using word2vec.

In this study, word2vec is applied to the data set of recipes. Word2vec can be applied by considering recipes as documents and ingredients as words. We do not include a window size parameter, since it is used to encode the ordering of words in document where it is relevant.  In recipes, the listing of ingredients is unordered. The objective function is 
\begin{equation}
\sum_{r \in R} \sum_{w_i \in r}  \sum_{j \neq i} \log P(w_{j}|w_i) \label{eq:02}
\end{equation}
where $R$ is a set of recipes, $r$ is a recipe, and $w_i$ is the $i$th ingredient in recipe $r$. The architecture is described in middle of Figure \ref{word2vec_arch}. The objective function is to maximize the likelihood of the prediction of the ingredient $w_j$ in the same recipe given the ingredient $w_i$. The probability is defined below.
\begin{equation}
P(w_j|w_i) = \frac{\exp(v_{w_i}^Tv_{w_j}^{'})}{\sum_{w \in W} \exp(v_{w_i}^Tv_w^{'})}
\end{equation}
where $w$ is an ingredient, $v_w \in \mathbb{R}^K$ is an input vector of ingredient, $v^{'}_w \in \mathbb{R}^K$ is an output vector of ingredient, $K$ is the dimension of the vector, and $W$ is the set of all ingredients.

Each ingredient is vectorized by word2vec, and the similarity of each ingredient is calculated using cosine similarity. Through vectorization in word2vec, those of the same genre are placed nearby. In other words, by using the word2vec vector, it is possible to select ingredients with similar genres. 

Next, we extend word2vec to be able to incorporate information of the country. When we vectorize the countries, we can calculate the analogy between countries and ingredients. For example, this method can tell us what is the French ingredient that corresponds to Japanese soy sauce by calculating ``Soy sauce - Japan + French = ?".

The detail of our method is as follows. We maximize objective function (\ref{eq:02}). 

\begin{equation}
\sum_{r \in R} \sum_{w_i \in r} \left(    \log P(w_{i}|c_r) +  \log P(c_r|w_{i}) + \sum_{j \neq i} \log P(w_{j}|w_i)\right)\label{eq:02}
\end{equation}
where $R$ is a set of recipes, $r$ is a recipe, $w_i$ is the $i$th ingredient in recipe $r$, and $c_r$ is the country recipe $r$ belongs to. The architecture is described in right of Figure \ref{word2vec_arch}. The objective function is to maximize the likelihood of the prediction of the ingredient $w_j$ in the same recipe given the ingredient $w_i$ along with the prediction of the the ingredients $w_i$ given the country $c_r$ and the prediction of the the country $c_r$ given the ingredient $w_i$. The probability is defined below.

\begin{equation}
P(b|a) = \frac{\exp(v_{a}^Tv_{b}^{'})}{\sum_{c \in W} \exp(v_{a}^Tv_c^{'})}
\end{equation}
where $a$ is a ingredient or country, $b,c$ are also, $v_a \in \mathbb{R}^K$ is an input vector of ingredient or country, $v^{'}_a \in \mathbb{R}^K$ is an output vector of ingredient or country, $K$ is the dimension of vector, and $W$ is the set of all ingredients and all countries.

We can use hierarchical softmax or negative sampling \cite{mikolov2013distributed} to maximize objective function (\ref{eq:02}) and find the vectors of ingredients and countries in the same vector space. 

\begin{figure}[h!]
\begin{center}
\includegraphics[width=15cm]{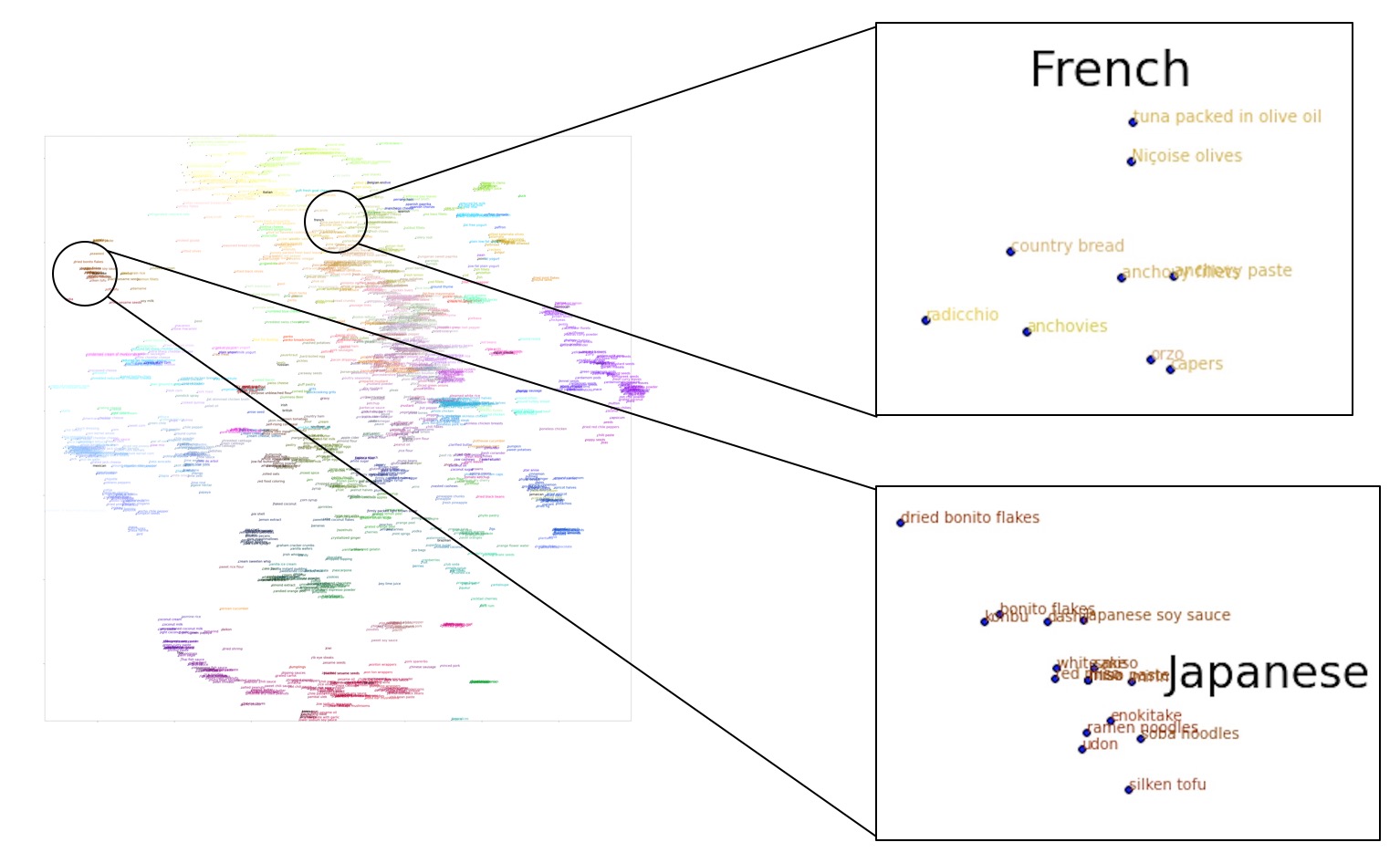}% This is a *.jpg file
\end{center}
\caption{Ingredients and countries map by extended word2vec: Each ingredient and country is mapped in 2D by using t-SNE. Also Each ingredient is colored by using t-SNE to convert 100 dimension vector into 3 dimension. The 3 dimension is corresponded to RGB color. Countries are represented by bold black.}\label{fig4}
\end{figure}

% For Original Research articles, please note that the Material and Methods section can be placed in any of the following ways: before Results, before Discussion or after Discussion.
Table \ref{table2} shows the ingredients around each country in the vector space, and which could be considered as most authentic for that regional cuisine \cite{ahn2011flavor}. Also, Figure \ref{fig4} shows the ingredients and countries in 2D map by using t-SNE method \cite{maaten2008visualizing}.

\begin{table}[htbp]
\begin{center}
\caption{Authentic ingredients for each country. Top 5 ingredients around each country in the vector space.}
\begin{tabular}{|c|c|c|c|c|}
\hline
 & French & Japanese & Italian & Mexican \\ \hline \hline
Top1 & Cognac & Mirin & Grated parmesan cheese & Corn tortillas\\ \hline
Top2  & Calvados & Dashi & pecorino romano cheese & Salsa\\ \hline
Top3  & Thyme springs & Nori & prosciutto & Tortilla chips\\ \hline
Top4  & Gruyere cheese & Wasabi paste &marinara sauce & Guacamole\\ \hline
Top5 & Nicoise olives & Bonito flakes &Sweet italian sausage& Poblano peppers \\ 
\hline
\end{tabular}
\label{table2}
\end{center}
\end{table}

\section{Experiment}
Our substitution strategy is as follows.
First we use extended word2vec and train it by Yummly dataset. After that all ingredients and countries are vectorized into 100 dimensional vector space. Second we find substitution by analogy calculation. For example, to find french substitution of Mirin, we calculate ``Mirin - Japanese + French" in the vector space and get the vector as result. After that we find similar ingredients around the vector by calculating the cosine similarity.

\begin{figure}[h!]
\begin{minipage}[b]{.25\linewidth}
\centering\includegraphics[width=3.4cm, angle=90]{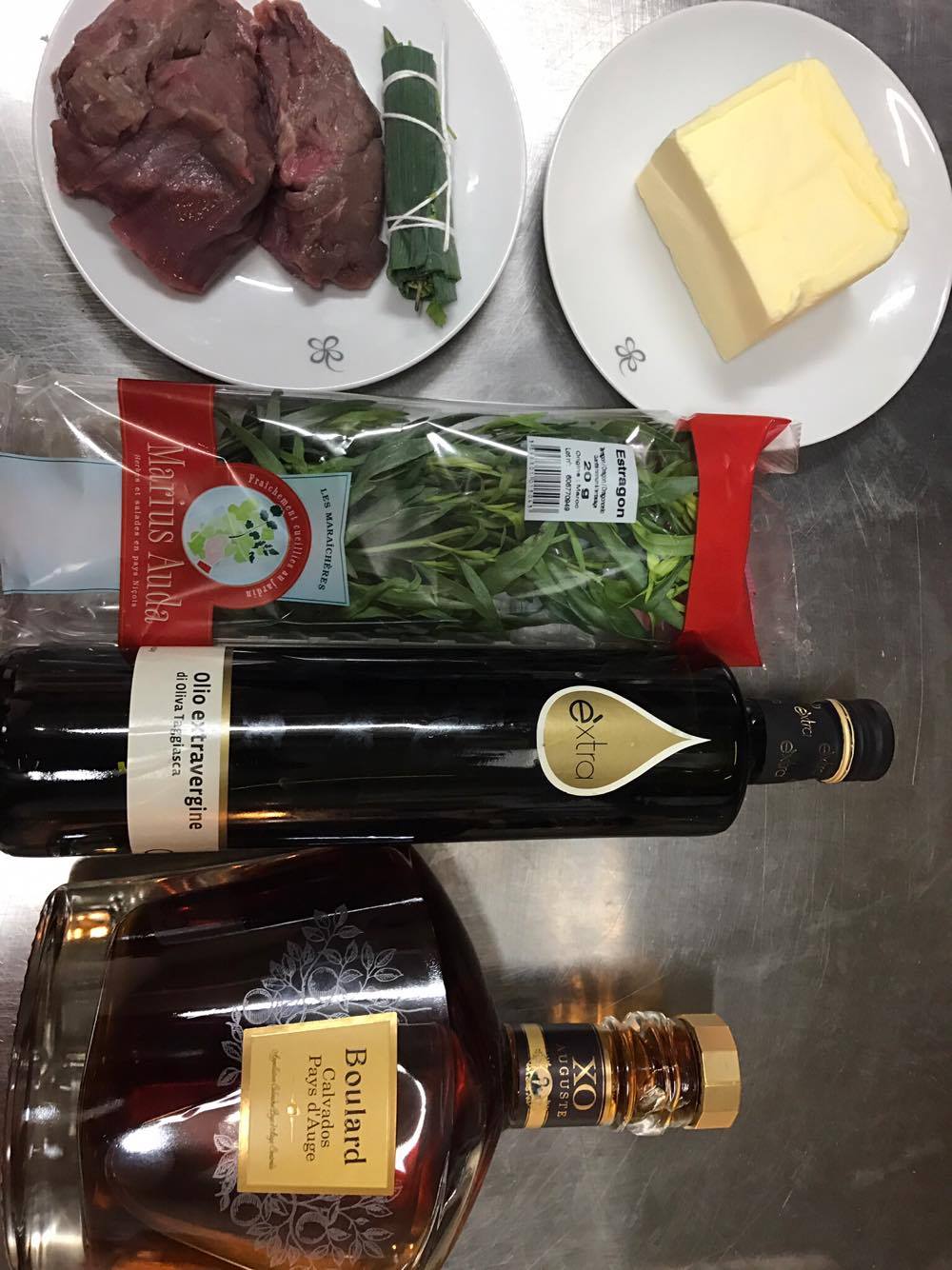}% This is a *.eps file
%\subcaption{Ingredients}\label{fig:2a}
\end{minipage}%
\begin{minipage}[b]{.25\linewidth}
\centering\includegraphics[width=3.4cm, angle=90]{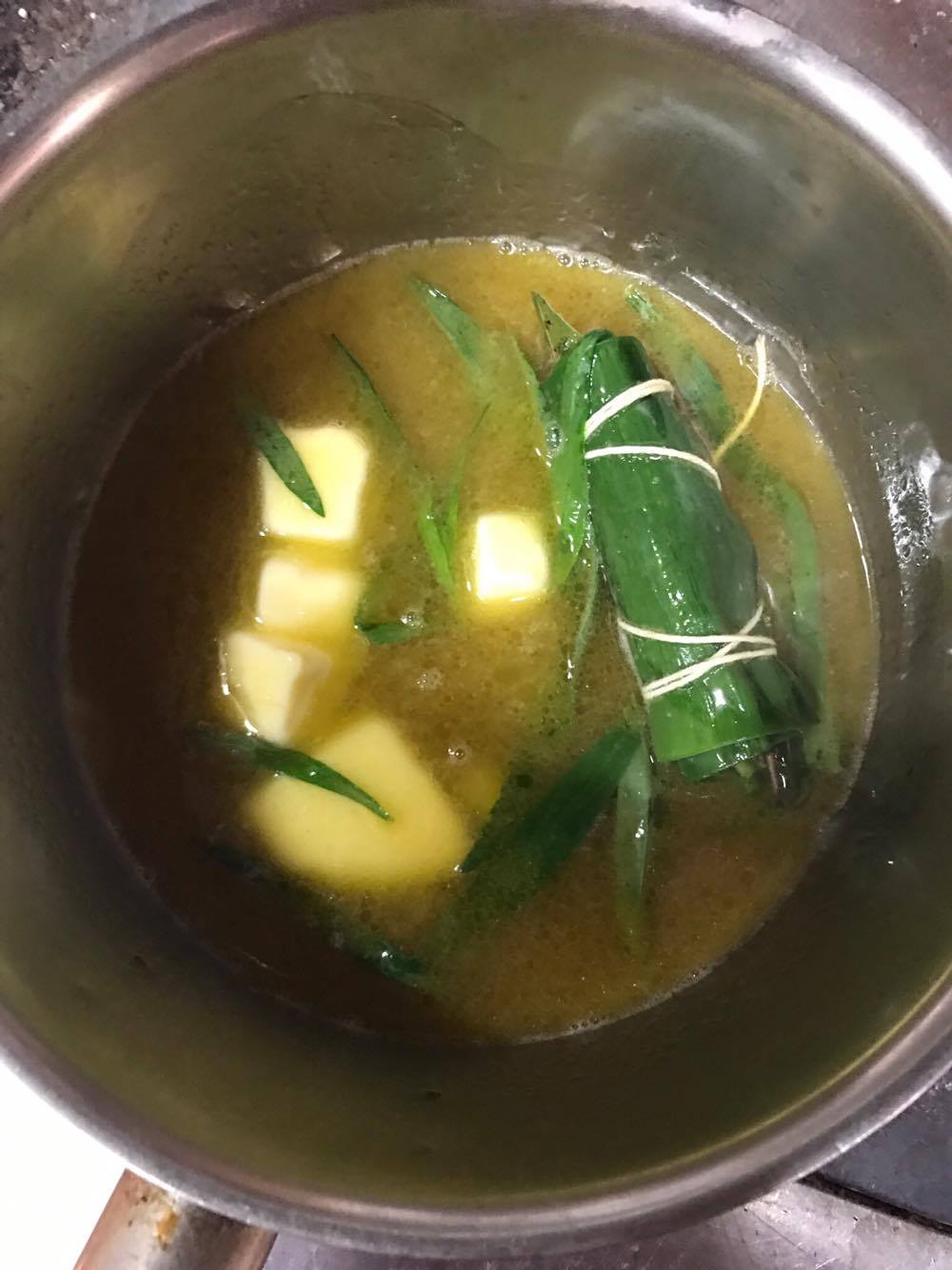}% This is an *.eps file
%\subcaption{Boil}\label{fig:2b}
\end{minipage}%
\begin{minipage}[b]{.25\linewidth}
\centering\includegraphics[width=3.4cm, angle=90]{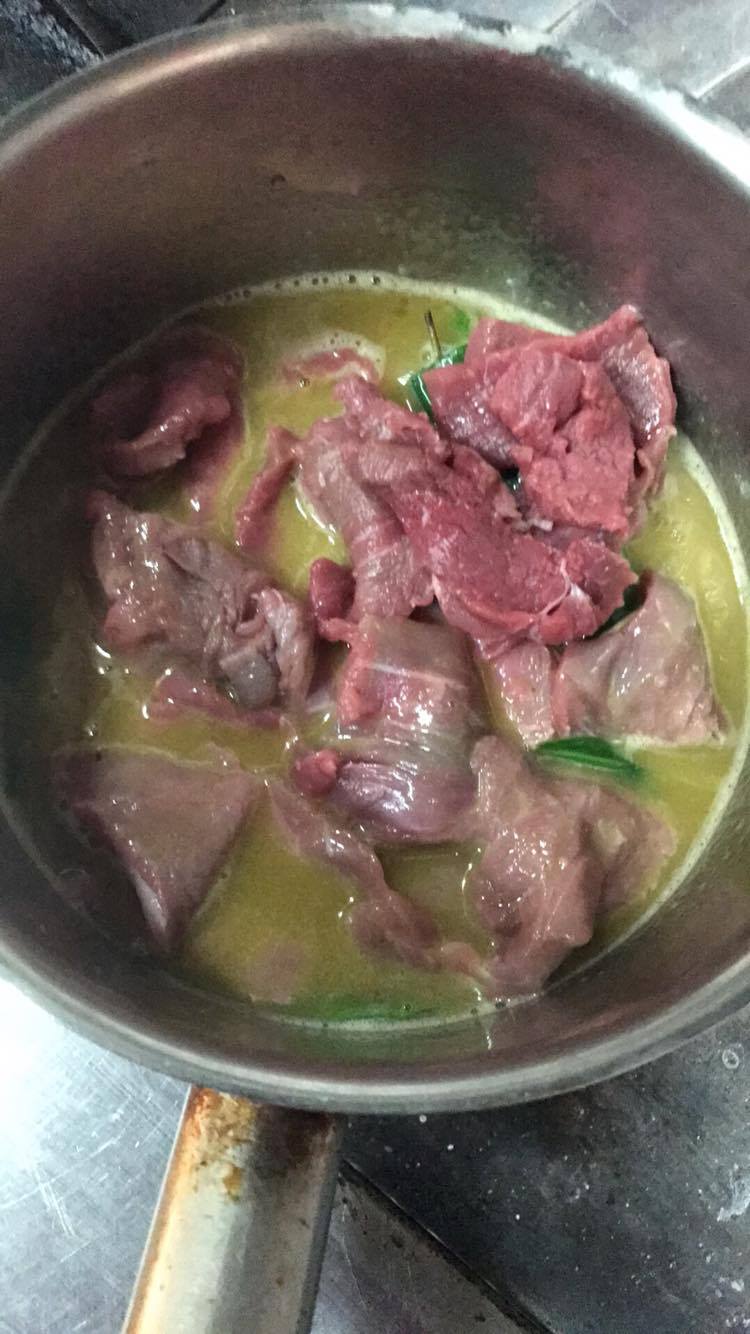}% This is an *.eps file
%\subcaption{Boil}\label{fig:2c}
\end{minipage}%
\begin{minipage}[b]{.25\linewidth}
\centering\includegraphics[width=3.4cm, angle=90]{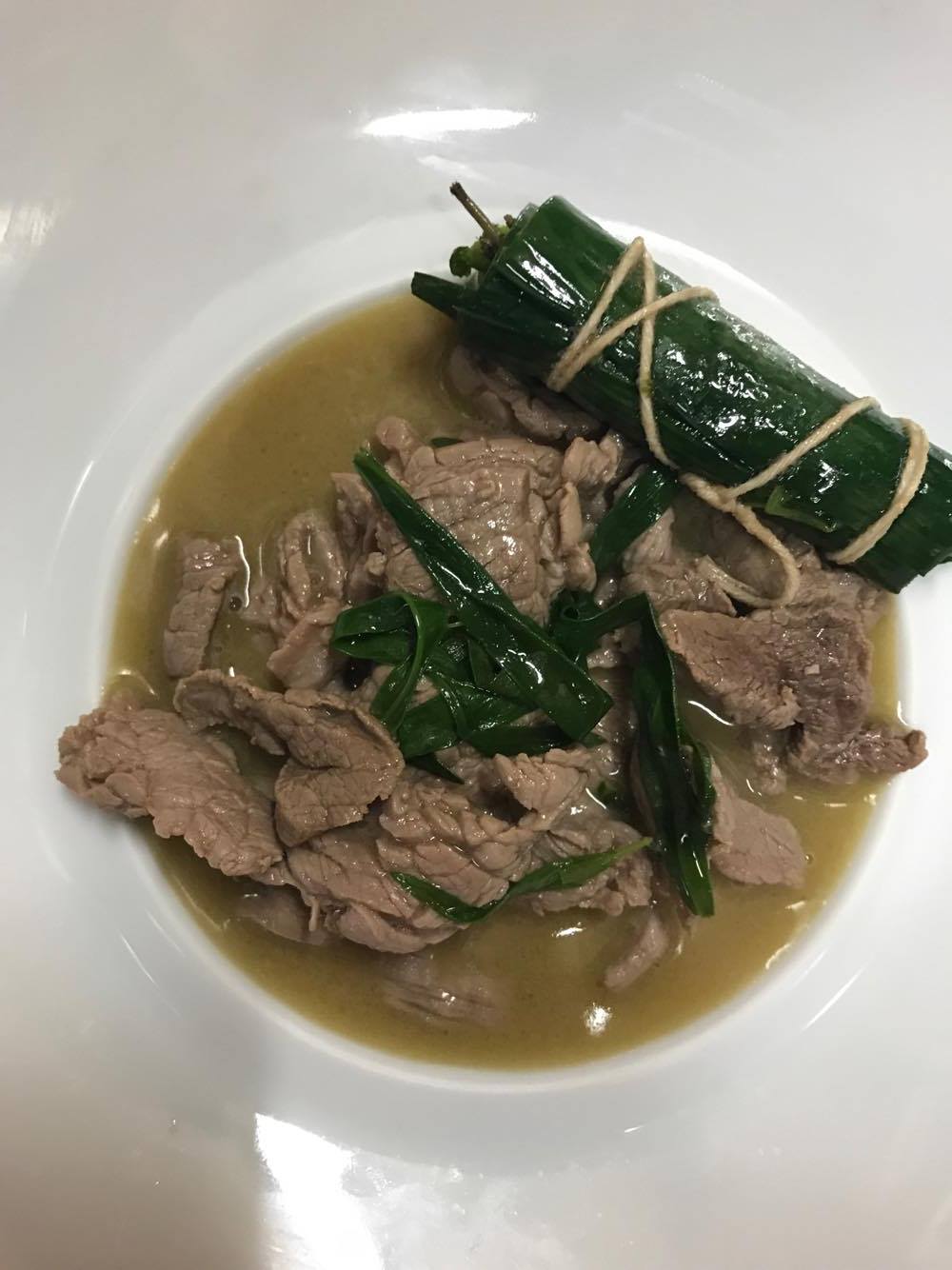}% This is an *.eps file
%\subcaption{Another subfigure}\label{fig:2d}
\end{minipage}%
\caption{Sukiyaki in French style. Professional chef KM who is one of the authors of this paper cooked the recipe suggested by our system.}\label{fig:sukiyaki_french}
\end{figure}

As an example of our proposed system, we transformed a traditional Japanese ``Sukiyaki" into French style. Table \ref{table3} shows the suggested replaceable ingredients and the probability after replacing. ``Sukiyaki" consists of soy sauce, beef sirloin, white sugar, green onions, mirin, shiitake, egg, vegetable oil, konnyaku, and chinese cabbage. Figure \ref{fig:sukiyaki_french} shows the Sukiyaki in French style cooked by professional chef KM who is one of the authors of this paper. He assesses the new recipe as valid and novel to him in terms of Sukiyaki in French. 
Here our task is in generating a new dish, for which by definition there is no ground truth for comparison.
Rating by experts is the standard approach for assessing novel generative artifacts, e.g. in studies of creativity \cite{jordanous2012standardised}, but going forward it is important to develop other approaches for assessment.

\begin{table}[htbp]
\begin{center}
\caption{Alternative ingredients suggested by extended word2vec model and country probability of changing food ingredients in order from the top. Professional chef KM who is one of the authors of this paper chose one alternative ingredient from top 10 suggested ingredients each.}
\begin{tabular}{|c|c|r|r|r|}
\hline
 Original Ingredient & Alternative Ingredient & P(Japanese) & P(French)&\# of replacement  \\ \hline \hline
-  & - & 0.974 & 0.000 & 0\\ \hline
Mirin  & Calvados & 0.552 & 0.009 & 1\\ \hline
Vegetable oil  & Olive oil & 0.393 & 0.031 & 2\\ \hline
Soy sauce  & Bouquet garni & 0.011 & 0.976 & 3\\ \hline
Green onions& Fresh tarragon & 0.000 &0.997 & 4 \\ \hline
Egg & Melted butter& 0.000 &0.999 & 5 \\ 
\hline
\end{tabular}
\label{table3}
\end{center}
\end{table}

\section{Discussion}
With growing diversity in personal food preference and regional cuisine style, the development of data-driven systems which can transform recipes into any given regional cuisine style might be of value for food companies or professional chefs to create new recipes.

In this regard, this study adds two important contributions to the literature. First, this is to the best of our knowledge, the first study to identify a recipe’s mixture of regional cuisine style from the large number of recipes around the world. Previous studies have focused on assessing degree of adherence to a single regional cuisine pattern. For example, Mediterranean Diet Score is one of the most popular diet scores. This method uses 11 main items (e.g., fruit, vegetable, olive oil, and wine) as criteria for assessing the degree of one’s Mediterranean style \cite{panagiotakos2006dietary}. However, in this era, it is becoming difficult to identify a recipe’s regional cuisine style with specific country/regional style. 
For example, should Fish Provencal, whose recipe name is suggestive of Southern France, be cast as French style? The answer is mixture of different country styles: 32\% French; 26\% Italian; and 38\% Spanish (see Figure \ref{fig3}).
%Specifically, based on our algorithm, the California Roll is the mixture of dietary style as follows: Japanese (XX%); Western American (XX%); and the rest (XX%). 

Furthermore, our identification algorithm can be used to assess the degree of personal regional cuisine style mixture, using the user’s daily eating pattern as inputs. For example, when one enters the recipes that one has eaten in the past week into the algorithm, the probability values of each country would be returned, which shows the mixture of regional cuisine style of one’s daily eating pattern. As such, a future research direction would be developing algorithms that can transform personal regional cuisine patterns to a healthier style by providing a series of recipes that are in accordance with one’s unique food preferences.

Our transformation algorithm can be improved by adding multiple datasets from around the world. Needless to say, lack of a comprehensive data sets makes it difficult to develop algorithms for transforming regional cuisine style. For example, Yummly, one of the largest recipe sites in the world, is less likely to contain recipes from non-Western regions. Furthermore, data on traditional regional cuisine patterns is usually described in its native language. As such, developing a way to integrate multiple data sets in multiple languages is required for future research. 

One of the methods to address this issue might be as follows: 1) generating the vector representation for each ingredient by using each data set independently; 2) translating only a small set of common ingredients among each data set, such as potato, tomato, and onion; 3) with a use of common ingredients, mapping each vector representation into one common vector space using a canonical correlation analysis \cite{kettenring1971canonical}, for example. 

Several fundamental limitations of the present study warrant mention. First of all, our identification and transformation algorithms depend on the quantity and quality of recipes included in the data. As such, future research using our proposed system should employ quality big recipe data.
Second, the evolution of regional cuisines prevents us from developing precise algorithm. For example, the definition of Mediterranean regional cuisine pattern has been revised to adapt to current dietary patterns \cite{serra2004does,kinouchi2008non}. Therefore, future research should employ time-trend recipe data to distinctively specify a recipe’s mixture of regional cuisine style and its date cf.~\cite{varshney2013flavor}. 
Third, we did not consider the cooking method (e.g., baking, boiling, and deep flying) as a characteristic of country/regional style. Each country/region has different ways of cooking ingredients and this is one of the important factors characterizing the food culture of each country/region. 
Fourth, the combination of ingredients was not considered as the way to represent country/regional style. For example, previous studies have shown that Western recipes and East Asian recipes are opposite in flavor compounds included in the ingredient pair \cite{zhu2013geography,varshney2013flavor,jain2015spices,tallab2016exploring,ahn2011flavor}. For example, Western cuisines tend to use ingredient pairs sharing many flavor compounds, while East Asian cuisines tend to avoid compound sharing ingredients. It is suggested that combination of flavor compounds was also elemental factor to characterize the food in each country/region. As such, if we analyze the recipes data using flavor compounds, we might get different results.

In conclusion, we proposed a novel system which can transform a given recipe into any selected regional cuisine style. This system has two characteristics: 1) the system can identify a degree of regional cuisine style mixture of any selected recipe and visualize such regional cuisine style mixture using a barycentric Newton diagram; 2) the system can suggest ingredient substitution through extended word2vec model, such that a recipe becomes more authentic for any selected regional cuisine style. Future research directions were also discussed.

\section*{Conflict of Interest Statement}
%All financial, commercial or other relationships that might be perceived by the academic community as representing a potential conflict of interest must be disclosed. If no such relationship exists, authors will be asked to confirm the following statement: 

The authors declare that they have no conflict of interest.

\section*{Author Contributions}
MK, LRV, and YI had the idea for the study and drafted the manuscript. MK performed the data collection and analysis. MS, CH, and KM participated in the interpretation of the results and discussions for manuscript writing and finalization. All authors read and approved the final manuscript.

\section*{Funding}
Varshney's work was supported in part by the IBM-Illinois Center for Cognitive
Computing Systems Research (C3SR), a research collaboration as part of the
IBM AI Horizons Network.

\section*{Acknowledgments}
This study used data from Yummly. We would like to express our deepest gratitude to everyone who participated in this services. We thank Kush Varshney for suggesting the spectral graph drawing approach to placing countries on the circle.

\bibliographystyle{IEEEtran}
\bibliography{sukiyaki}

\end{document}